\documentclass[prl,nobalancelastpage,superscriptaddress,twocolumn]{revtex4}
\usepackage{graphicx,amssymb}

\newcommand{\gate}[1]{\textsc{#1}}

\newlength{\placeimageheight}

\newcommand{\placeimage}[2]{
\setlength{\placeimageheight}{#1}
\raisebox{0.5ex}{\raisebox{-0.5\placeimageheight}{
	\includegraphics[height=\placeimageheight]{#2}}}
}

\begin{document}

\title{Programmable networks for quantum algorithms}
\author{Norbert Schuch}
\affiliation{Institut f\"ur Theoretische Physik, Universit\"at
Regensburg, D--93040 Regensburg, Germany}
\affiliation{Max-Planck-Institut f\"ur Quantenoptik,
Hans-Kopfermann-Str.~1, D--85748 Garching, Germany}
\author{Jens Siewert}
\affiliation{Institut f\"ur Theoretische Physik, Universit\"at
Regensburg, D--93040 Regensburg, Germany}

\begin{abstract}
The implementation of a quantum computer requires the realization
of a large number of $N$-qubit unitary operations which represent
the possible oracles or which are part of the quantum algorithm.
Until now there are no standard ways to uniformly generate whole
classes of $N$-qubit gates.
We have developed a method to generate arbitrary controlled phase 
shift operations with a \emph{single} network of one-qubit and two-qubit
operations. This kind of network can be adapted to various
physical implementations of quantum computing and is suitable
to realize the Deutsch--Jozsa algorithm as well as Grover's
search algorithm.
\end{abstract}

\maketitle

The experimental 
implementation of complex $N$-qubit operations (where $N\ge 3$)
and the realization of complete quantum algorithms are 
major challenges in the field of quantum computation.
Any progress in this direction proves the practical
feasibility of quantum computation. 
Further, it provides a tool
for systematic study of the physical und technological requirements 
for quantum computers
such as parametric constraints of a given implementation, decoherence
times, robustness of many-qubit entanglement, 
measurement efficiency etc.

The attempts of practical implementation focus essentially
on Shor's algorithm~\cite{shor94}, Grover's 
database search~\cite{grover97},
and the Deutsch--Jozsa algorithm 
\cite{deutsch85,deutsch92}. 
Qubit-based experimental realizations of quantum algorithms 
have been achieved
with liquid-state NMR techniques \cite{jones98,vandersypen00,%
collins00,kim00,vandersypen01} and with trapped ions~\cite{gulde03}.
In order to practically implement (or rather simulate) a quantum
computer one needs to complete the following steps:
prepare the initial state, apply an $N$-bit unitary
operation which encodes the properties of a certain function $f$ (the
so-called {\em oracle}), 
perform another sequence of operations (the quantum algorithm 
that extracts the properties of $f$), 
and finally measure the qubit register which contains the desired
information about $f$.

The universality of quantum computation implies that it
is possible to generate arbitrary $N$-qubit gates
by using sequences of one-qubit and two-qubit 
operations only~\cite{deutsch95-prsla-univ,lloyd95,bremner02}. 
Barenco {\em et al.}, and later Cleve {\em et al.}, have developed
methods to design networks for $N$-qubit controlled 
operations~\cite{barenco95-pra,cleve:qalg_revisited}.
While these methods, in principle, are sufficient to generate 
any $N$-qubit gate  required for the known quantum algorithms,
it is not obvious how to apply them 
for a systematic and efficient {\em practical} realization
of the operations. 

A feature common to all existing implementations so far is that 
the sequences of one-qubit and two-qubit operations to realize 
the algorithm depend on the specific physical system and the number of qubits.
More importantly, they even depend on the particular choice among 
the possible functions $f$. 
For example, all $N=3$ implementations of the Deutsch--Jozsa
algorithm \cite{collins00,kim00,siewert01}
use some classification of the 
$(2^N)!/((2^{N-1})!)^2$ balanced functions and give prescriptions
how to realize the functions in each class. While for $N=3$ there are
only 70 balanced functions, this approach appears hard to extend 
even to $N=4$.
This situation is not satisfactory.
Clearly, scalability is a requirement not only for quantum
hardware~\cite{divincenzo00}, but also for quantum software.

In practical realizations of quantum information processing,
one has to cope also with other problems: firstly,
it is often not possible to directly perform
two-qubit operations between arbitrary pairs of bits. Secondly,
the controlled-\gate{not} (\gate{cnot}) 
gate is not the genuine two-qubit gate for many
proposed qubit systems.
Different interaction Hamiltonians provide different
types of two-qubit gates in a ``natural way'', i.e., gates
which can be achieved 
with a single two-qubit operation~\cite{makhlin00,vidal02,schuch03}.

In order to overcome these difficulties, we 
have worked out a systematic approach to design quantum networks 
which are suitable to perform arbitrary controlled
phase-shift operations 
\begin{equation}
   U_{\vec{\theta}}:\ \
   | \mathbf x \rangle\ \longrightarrow\ 
     \mathrm e^{-i\theta_{\mathbf x}}
   | \mathbf x \rangle\ 
\end{equation}
on $N$ qubits 
(here $| \mathbf x \rangle=|x_1,\ldots,x_N\rangle, 
x_j \in \{0,1\}$, denotes an element of the
$N$-qubit computational basis). The most important
feature of these networks is that  the parameters 
$\theta_{\mathbf x}$
of the $N$-bit operations are determined {\em exclusively}
by the rotation angles of single-qubit operations.
Therefore, these networks may be regarded as
programmable (in a technical sense): one
can imagine to load network instructions into
the processor with a punch tape where the
instructions are encoded by the
$2^N-1$ rotation angles.
This scheme resembles the structure of information
processing in conventional computer processors and 
may be considered as a step 
towards a von Neumann Architecture for quantum computers.

We show that this kind of network can be extended recursively
to an arbitrary
number of qubits assuming only nearest-neighbor coupling.
As the method utilizes only one type of two-qubit gate,
it can be, in principle, adapted to a large class of
hardware implementations. As an example we discuss
the implementation of the network with up to four
Josephson charge qubits. 
\pagebreak

The elementary one-qubit gate in our networks is the
phase shift ($z$ rotation)
$$
\placeimage{1.75em}{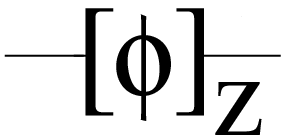}=
\left(\begin{array}{cc}e^{-i\phi/2}&\\&e^{i\phi/2}\end{array}\right)\quad.
$$
Note that \hspace{-1em}\placeimage{1.75em}{phi_z.eps}
does not mix the states $\{|0\rangle,|1\rangle\}$ of
the computational basis, but generates a phase shift
whose sign depends on the qubit state.
The fundamental two-qubit gate is the \gate{cnot}
$$
\mbox{\placeimage{4em}{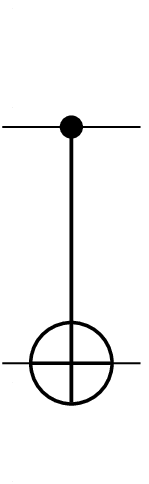}}=
\left(\begin{array}{cccc}
1\\&1\\&&0&1\\&&1&0
\end{array}\right)\quad.
$$

\begin{figure}[b]
\placeimage{5em}{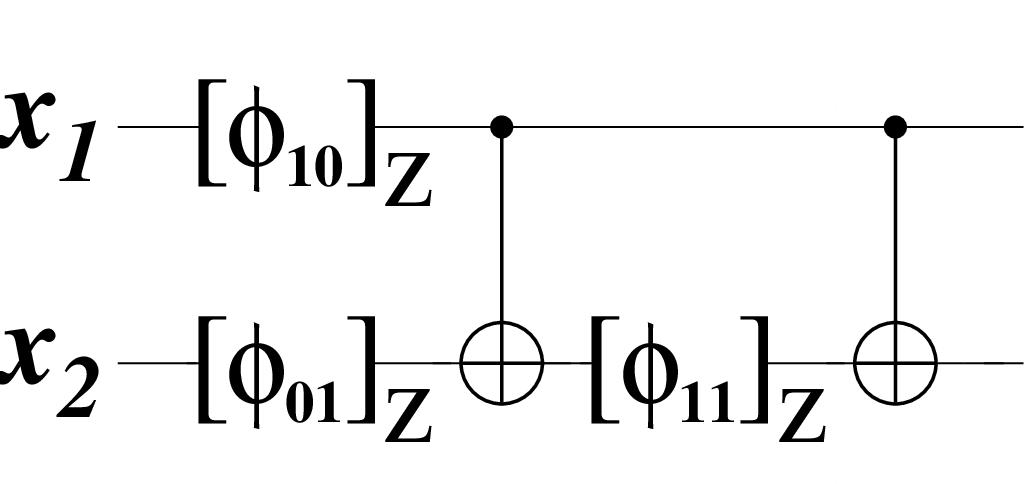}
\caption{%
Phase-shifting network for two qubits.
The first two $z$
      rotations change only the phase of
      the input state $|x_1,x_2\rangle$:
      the gate \protect\hspace{-1em}
               \protect\placeimage{1.75em}{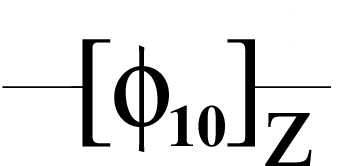}
      on the $x_1$ line shifts the phase of the input state 
      by $-\phi_{10}/2$ if $x_1=0$ 
      or by $+\phi_{10}/2$ if $x_1=1$. 
      The gate \protect\hspace{-1em}
          \protect\placeimage{1.75em}{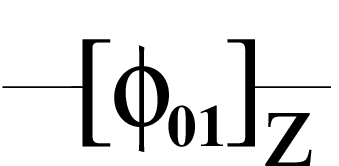} acts
       correspondingly on $x_2$. 
      The first \gate{cnot} gate generates the state 
      $|x_1,x_1\oplus x_2\rangle$ without further changing the phase.
      The line $x_2$ contains now the result of $x_1\oplus x_2$ 
      (where $\oplus$ is the addition modulo $2$).
      Subsequent application of
      \protect\hspace{-1em}
      \protect\placeimage{1.75em}{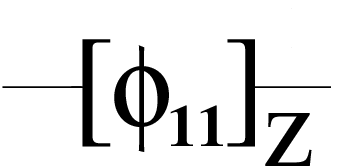} on this line 
      results in a {\em conditional} shift of the
       phase by $-\phi_{11}/2$ if $x_1\oplus x_2=0$ or by $\phi_{11}/2$
      if $x_1\oplus x_2=1$. The second \gate{cnot} restores the state 
      $|x_1,x_2\rangle$ so that 
      the input state is left unchanged apart from a phase shift that
      depends on the value of $x_1$, $x_2$ and $x_1\oplus x_2$.} 
\label{fig:2bit_straight}
\end{figure}
To motivate the discussion of the general method below,
let us consider the simple two-bit network in 
Fig.~\ref{fig:2bit_straight}.
The easiest way to understand this network 
is to analyze its action on the states of the computational basis 
$|\mathbf x\rangle=|x_1,x_2\rangle$.
We see that the network leaves the basis states unchanged, it merely
generates a phase factor that depends on the index $\mathbf x$ 
of the basis state 
(see the caption of Fig.~\ref{fig:2bit_straight}).
Therefore, the action on a superposition of basis states is that
the modulus of the amplitude for each component remains the same 
while the relative phases change in a well-defined way.

The total phase shift $\theta_{\mathbf x}$ 
due to the action of the network
$
|\mathbf x\rangle\mapsto e^{-i\theta_\mathbf x}|\mathbf x\rangle
$
can be written more formally as 
\begin{eqnarray}
\theta_\mathbf x & = & (-1)^0\ \frac{\phi_{00}}{2}
                   \   +\ (-1)^{1\cdot x_1\oplus 0\cdot x_2}\ \frac{\phi_{10}}{2}+
       \nonumber \\
                 && 
       + (-1)^{0\cdot x_1\oplus1\cdot x_2}\ \frac{\phi_{01}}{2} +
	 (-1)^{1\cdot x_1\oplus 1\cdot x_2}\ \frac{\phi_{11}}{2}\ .
\label{eq:theta_2bit}
\end{eqnarray}
Here we have added a 
(physically irrelevant) global phase
$\phi_{00}/2$ which applies uniformly to all basis states.

By introducing the inner product modulo two of the binary
strings 
$\mathbf x= (x_1,\ldots,x_N)$ and
$\mathbf y= (y_1,\ldots,y_N)$ ($x_j,y_j \in\{0,1\}$) as
$\mathbf x\cdot\mathbf y:= 
           x_1 \cdot y_1\oplus\ldots \oplus x_N\cdot  y_N$
we see that the
phase shift $\theta_{\mathbf x}$ according to
Eq.~(\ref{eq:theta_2bit}) is equal to
\begin{equation}
\theta_\mathbf x\ =\ \frac{1}{2}\ \sum_{\mathbf y=0}^{2^2-1}\ 
                   (-1)^{\mathbf x\cdot\mathbf y}
                    \phi_{\mathbf y}\quad.
\label{hadamard2}
\end{equation}
That is, the phase shifts 
$\vec{\theta}=(\theta_{00},\ldots,\theta_{11})$
are related to the one-qubit rotation angles $\vec{\phi}=
(\phi_{00},\ldots,\phi_{11})$ by an (unnormalized) 
two-qubit Hadamard transformation $\mathcal H$
(the matrix of this transformation has
the entries 
$
\mathcal H_{\mathbf x,\mathbf y}=(-1)^{\mathbf x\cdot\mathbf y}\ ).
$
Each term in the sum of Eq.~(\ref{hadamard2}) represents a
conditional phase shift by an angle $\phi_{\mathbf y}$
where the binary representation of the index $\mathbf y$ 
indicates the digits $x_j$ that are part of the 
corresponding \gate{xor} control condition.

Thus we can summarize that the two-qubit network 
acts on states of the computational basis by shifting their
phase. The total phase shift is obtained by subsequently
applying all possible conditional phase shifts which can
be derived by combining the digits $x_j$ of the input state.
The conditional phase shifts are realized by applying
a single-qubit rotation to the
qubit that contains the result of the corresponding control
condition.

It is evident that this scheme can be generalized to
an arbitrary number $N$ of qubits:\\
Provided that we are able to construct an $N$-qubit network
which, on application to any state
$|\mathbf x\rangle$ of the $N$-qubit computational basis,
\begin{itemize}
   \item[\textbf{[N1]}] allows to generate---following {\em classical}
         logics---all possible control conditions
         $\mathbf{x\cdot y}$ from the digits of $\mathbf x$ (where
         $\mathbf y$ is a binary string of length $N$),
   \item[\textbf{[N2]}] applies $z$ rotations to each condition
         $\mathbf{x\cdot y}$, 
\end{itemize}
we can implement any generalized phase shift operator 
$U_{\vec{\theta}}: 
|\mathbf x\rangle\mapsto e^{-i\theta_\mathbf x}|\mathbf x\rangle$
by using one and the same network, merely by adjusting 
the angles $\vec{\phi}=(\phi_0,\ldots,\phi_{2^N-1})$
of one-qubit $z$ rotations. By virtue of 
$\mathcal H^2=2^N\openone$,
the angles 
$\vec{\phi}$ are related to the desired phases $\vec{\theta}=
(\theta_0,\ldots,\theta_{2^N-1})$ simply by an
$N$-bit Hadamard transformation
\begin{equation}
\vec\phi=\frac{1}{2^{N-1}}\mathcal H\vec\theta\quad.
\label{nhadamard}
\end{equation}
The network can be viewed as a ``black box'' with $2^N-1$ ``knobs'' 
whose settings determine the gate $U_{\vec\theta}$ 
realized by the black box. By choosing $\vec{\phi}$, any phase configuration  
$\vec\theta$ can be programmed into the network.

\setcounter{figure}{2} 
\begin{figure*}[t]
\includegraphics[width=\textwidth]{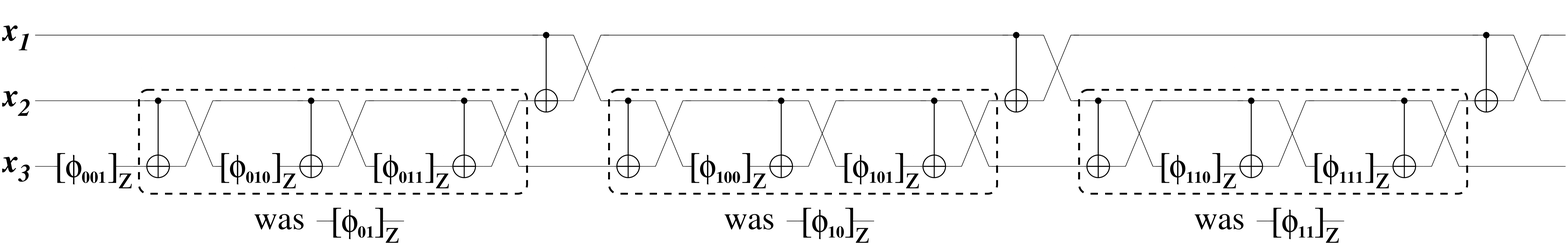}
\caption{\label{fig:syst-3-network}
The three-bit network for a system with nearest-neighbor $XY$ 
coupling. The dashed lines indicate the parts of the two-bit
network which have been replaced according to rule (ii).}
\end{figure*}
\setcounter{figure}{1}

The remaining task is to show that it is indeed possible to construct
such programmable networks 
for an arbitrary number of qubits.  
Since for many physical implementations 
coupling between arbitrary qubits is not available 
or falls off with increasing distance or qubit number,
we will assume only nearest-neighbor coupling between the qubits%
~\footnote{If we assume that a hardware setup is available
which allows to perform the \gate{cnot} operation between arbitrary
qubits, the networks can be constucted efficiently by ordering the
$\phi_{\mathbf y}$'s such that the $\mathbf y$'s form a Gray sequence
(see Ref.~\cite{barenco95-pra}).}.
Further, we will show that efficient network design is possible 
even if \gate{cnot} is not the natural two-qubit 
gate for a given implementation~\cite{makhlin00,vidal02,schuch03}.

\begin{figure}[b]
\placeimage{9em}{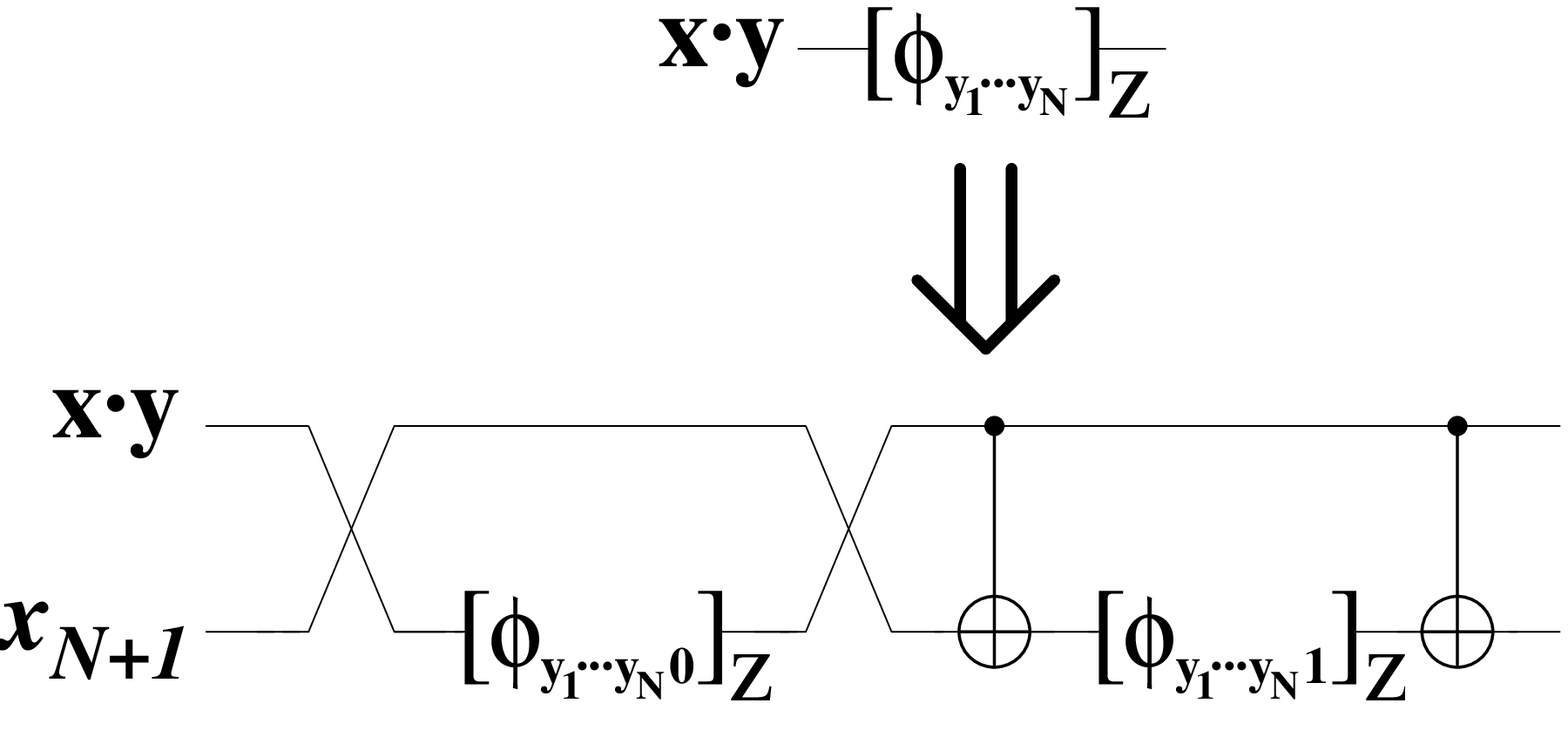}
\caption{Recursive extension of the network.
         By a \gate{swap} with the $N$+1st qubit
         the phase shift 
         \protect$\phi_{\mathbf{y_1\ldots y_N}}\rightarrow
                  \phi_{\mathbf{y_1\ldots y_N0}}$
         (controlled by the original condition) appears now on
the $N$+1st qubit. A second \gate{swap} and a subsequent
\gate{cnot} from the $N$-th to the $N$+1st qubit give
the original condition \gate{xor}ed with $x_{N+1}$---this
condition controls the second $z$ rotation.
We mention the identity\mbox{
               \protect\hspace{-1em}
               \protect\placeimage{2.35em}{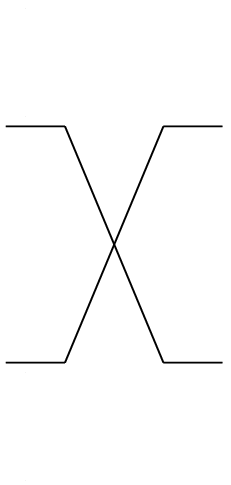}=
               \protect\hspace{-1em}
               \protect\placeimage{2.35em}{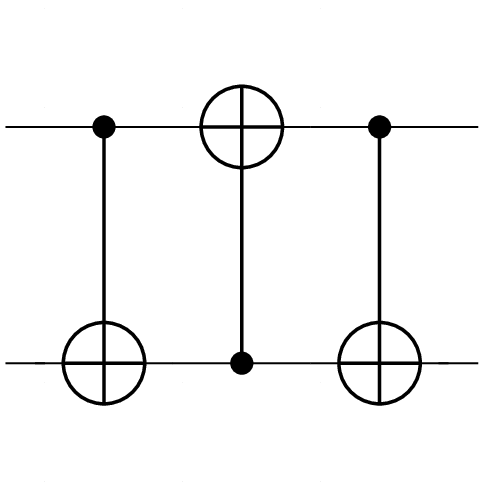}.}
        }
\label{fig_recursive}
\end{figure}

For the recursive 
method it is convenient that all 
controlled phase shifts appear on the last (i.e., the $N$-th)
qubit. Then the network
for $N+1$ qubits can be obtained from the network for $N$ bits simply  
by (i) adding the $N$+1st line with a $z$ rotation (i.e., a
phase shift controlled by $x_{N+1}$), and  (ii) by 
replacing each conditional $z$ rotation on the $N$-th
qubit in the $N$-bit network
by two rotations of the $N$+1st qubit, as shown 
in Fig.~\ref{fig_recursive}.
By following these rules, we get $z$ rotations for all possible \gate{xor} 
conditions of the $N+1$ qubits.  The conditions for 
the rotations characterized  by  the indices 
$\mathbf y=(y_1,\ldots,y_{N+1})$ appear
in their natural order if read as a binary code.

To illustrate this, we explain the method for up to three qubits with an
$XY$ interaction as, e.g., for Josephson charge qubits
coupled by SQUID loops~\cite{siewert01,echternach01}.
The construction is analogous for other two-bit
operations used as the basic element. 
The $XY$ interaction leads to a particularly compact 
design of the circuit. 
This is because the natural gate for this interaction
is again a classical gate, namely the product of 
\gate{cnot} and \gate{swap} denoted by \gate{cns}~\cite{schuch03}
$$
\gate{cns}\equiv\placeimage{5em}{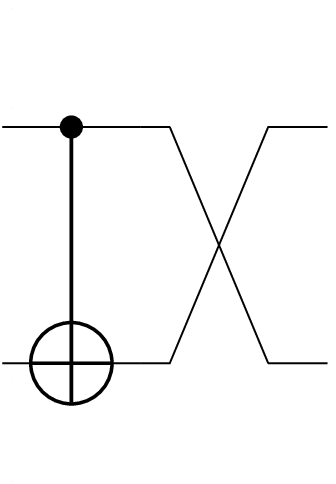}\quad.
$$
The ``one-qubit network'' is just a single $z$ rotation
\placeimage{1.75em}{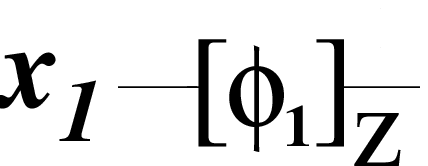}.
For two qubits we modify the network in 
Fig.~\ref{fig:2bit_straight} such that all
$z$ rotations appear on the second line
$$
\placeimage{5em}{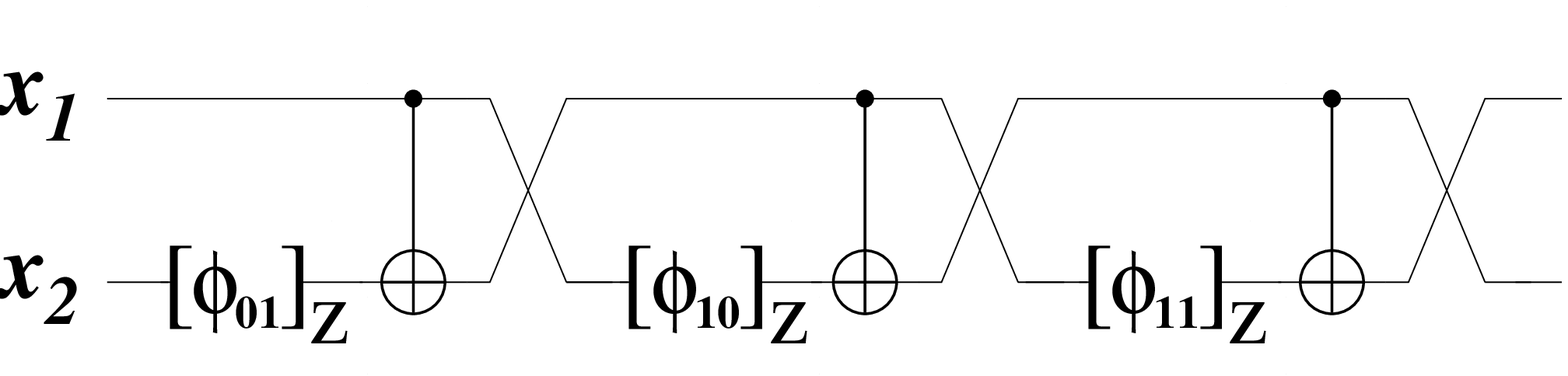}\quad.
$$
From here, the derivation of the three-qubit network 
is straightforward. The result is shown in Fig.~\ref{fig:syst-3-network}.

We emphasize that the central issue is the
concept of these networks expressed in the requirements [N1], [N2].
Since [N1] refers only to classical logics, optimized circuits
can be found, e.g., by an exhaustive search (on a classical computer).
An example is the  four-qubit network 
in Fig.~\ref{fig:par-4-network}. 

By realizing such a network with
a physical system (and supplementing it with 
one-qubit operations such as, e.g., the Hadamard gate), 
both the Deutsch--Jozsa algorithm and Grover's search algorithm
can be implemented.
The refined version of the Deutsch--Jozsa algorithm~%
\cite{cleve:qalg_revisited,collins98} requires 
the gate $U_f: |\mathbf x\rangle\mapsto
e^{-i\pi f(\mathbf x)}|\mathbf x\rangle$ where $f$ is a constant or
 balanced Boolean function.
Analogously, for Grover's algorithm 
the gate $U_g$ 
has to be implemented with a Boolean function $g$: 
$g(\mathbf x^{\ast})=1$ for one particular $\mathbf x^{\ast}$ 
and $g(\mathbf x\neq \mathbf{x^{\ast}})=0$. 
For the generalized Grover algorithm~\cite{boyer98}, several items may be
marked.  Hence the required phase
shifts $\theta_{\mathbf x}$ are given by $\pi f(\mathbf x)$
or $\pi g(\mathbf x)$, respectively. The corresponding one-bit rotation
angles $\vec{\phi}$ are readily obtained by  Hadamard-transforming
$\vec{\theta}$ according to Eq.\ (\ref{nhadamard}).

\setcounter{figure}{3}
\begin{figure*}[t]
\includegraphics[width=\textwidth]{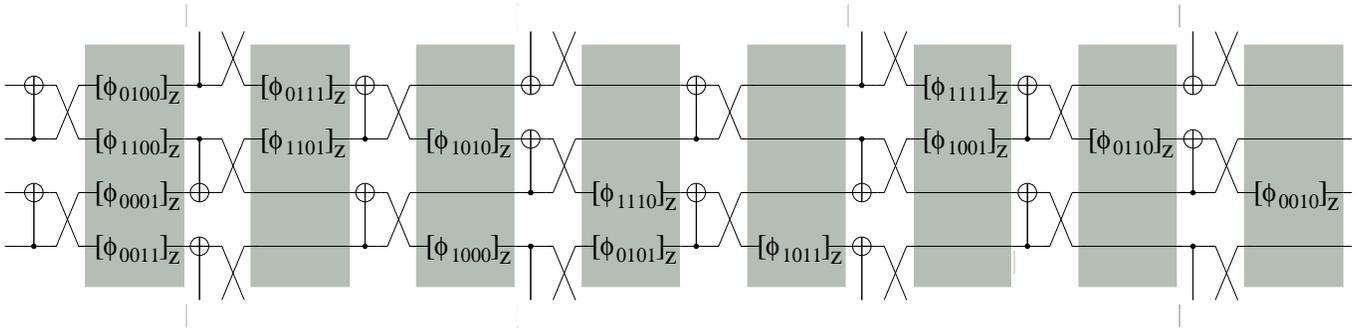}
\caption{\label{fig:par-4-network}
Optimized four-bit network for qubits with nearest-neighbor 
$XY$ coupling. With this network (plus the Hadamard gate for 
each qubit), 
both the Deutsch--Jozsa algorithm and database 
search can be realized, e.g., with Josephson charge qubits.
We have assumed qubits coupled in 
a chain with periodic boundary conditions. 
A particularly interesting feature
of this network is that nearest-neighbor coupling allows
for parallel execution of operations.
Note that the $z$ rotations and the two-qubit operations
appear in well-separated blocks.
The optimization criterion here was to minimize
the number of two-qubit blocks.}
\end{figure*}

We mention that there are other interesting applications 
for these programmable networks.
For example, by Hadamard-transforming the $N$-th
qubit a network is obtained which is capable
of generating a generalized controlled-\gate{not} operation
on the $N$-th qubit with an arbitrary control condition composed
from the other $N-1$ qubits.
Special cases are the Toffoli gate 
 (with an arbitrary number of control bits~\cite{barenco95-pra}) 
and the \gate{carry} operation  that appears in 
the network for the quantum adder used, e.g., in Shor's factoring 
algorithm~\cite{vedral96}. 

Thus, the new type of network presented here offers efficient practical
solutions for a wide range of computational tasks; this is
comprehensible for modest qubit numbers (on the order of 10) just
by comparing with existing solutions. 
On the other hand, the limit of
large $N$ requires more careful investigation. Although the 
parallel execution of operations (as shown in Fig.~\ref{fig:par-4-network}) 
can reduce, in principle, the complexity to $O(2^N/N)$ it is not clear
whether further substantial improvement is possible. There is
exponential scaling also for another resource. 
According to Eq.~(\ref{nhadamard}) the 
accuracy required for the one-qubit $z$ rotations will scale $\sim 2^{-N}$.
It remains subject for future work whether this
limit can be softened, e.g., for certain subclasses
of gates $U_{\vec\theta}$.

{\em Acknowledgements} -- The authors would like to thank
                       R.\ Fazio, E.\ Kashefi, Yu.\ Makhlin, K.\ Richter,
                       and V.\ Vedral
                       for stimulating discussions and comments.

\end{document}